\documentclass[aps,prd,twocolumn,nofootinbib,showpacs,preprintnumbers,floatfix,longbibliography]{revtex4-2}

\usepackage{natbib}
\usepackage{longtable}
\usepackage[pdftex]{graphicx}
\usepackage[usenames]{color}
\usepackage[dvips]{epsfig}

\usepackage{hyperref}
\usepackage{bbm}
\usepackage{marvosym}
\usepackage{verbatim}
\usepackage{psfrag}
\usepackage{slashed}

\usepackage{amsmath}
\usepackage{amssymb}
\usepackage{wasysym}
\usepackage{multirow}
\usepackage{subfigure}
\usepackage{comment}
\usepackage{kotex}
\usepackage{diagbox}
\usepackage{siunitx} 
\usepackage{array} 
\usepackage{amssymb}  
\usepackage{booktabs,amsmath}
\usepackage{bbold}
\usepackage[table,x11names]{xcolor}
\definecolor{maroon}{cmyk}{0,0.87,0.68,0.32}
\usepackage{amsfonts}

\definecolor{boxcolor}{HTML}{ffe6a1}

\DeclareMathAlphabet{\mathpzc}{OT1}{pzc}{m} {it}
\begin{document}
\title{Hybrid Classical-Quantum Sampling for Lattice Scalar Field Theory}

\author{Hee-Cheol Kim$^{\rm a}$}
\email{heecheol1@gmail.com}
\author{Jangho Kim$^{\rm b,c}$}
\email{kim.jangho1120@gmail.com}
\affiliation{$^{\rm a}$ Department of Physics, POSTECH, Pohang 37673, Korea}
\affiliation{$^{\rm b}$ Lattice Gauge Theory Research Center, Department of Physics and Astronomy, Seoul National University, Seoul 08826, Korea}
\affiliation{$^{\rm c}$ Institute for Advanced Simulation (IAS-4) \& JARA-HPC, Forschungszentrum J\"ulich, 52428 J\"ulich , Germany}

\begin{abstract}
We investigate lattice scalar field theory in two-dimensional Euclidean space via a quantum annealer. To accommodate the quartic interaction terms, we introduce three schemes for rewriting them as quadratic polynomials through the use of auxiliary qubits. These methods are applied on D-Wave quantum annealer, and their effectiveness is assessed by examining the annealer-generated distributions. Using these distributions, we perform Monte Carlo sampling via the Metropolis-Hastings algorithm and compare the outcomes with those from classical Metropolis simulations.
\end{abstract}

\maketitle

\section{Introduction}
Quantum computing has seen remarkable progress in recent years, both in terms of theoretical development and experimental implementation. Among the various architectures, quantum annealers stand out as some of the first commercially available quantum devices. These systems, designed to find ground states of Ising-like Hamiltonians via adiabatic evolution, now feature over 5000 physical qubits~\cite{dwave_pegasus_topology}, enabling exploration of large-scale optimization problems and beyond.

Recent studies have demonstrated that quantum annealers can be used not only for discrete optimization tasks but also for sampling problems, including those relevant to path integral formulations in quantum field theory. In particular, importance sampling techniques applied on quantum annealers have shown promising results for systems with integer-valued degrees of freedom~\cite{Kim:2023sie,Kim:2023pjk,Kim:2024wnb}. Quantum annealing has also been explored in contexts such as boson sampling~\cite{Lund:2017cfv}, protein folding~\cite{Ghamari:2024iat}, noisy Gibbs sampling~\cite{Vuffray:2020hjh}, hybrid quantum-classical methods~\cite{Jattana:2024otq}, thermal sampling~\cite{Izquierdo:2020acy}, and studies of phase transitions~\cite{Wild:2021ene}. Additionally, postprocessing techniques are often needed to recover the correct equilibrium distribution from raw annealer samples~\cite{Sandt:2023ewm, Ghamari:2022jyc, Shibukawa:2023vke}, and embedding challenges on the hardware graph can limit scalability~\cite{Quinton:2024rkl,Weinberg:2020mba}.

These findings raise the question of whether such approaches can be generalized to field theories involving continuous, real-valued variables—a necessary step toward simulating physically relevant models such as scalar field theories.

Scalar field theory (see, e.g.,\cite{De:2005ny}) offers a minimal yet robust framework for probing the dynamics of quantum fields. Earlier realizations of scalar field simulations on quantum platforms have demonstrated both real-time and Euclidean evolutions using quantum annealers and gate-based quantum computers \cite{Zemlevskiy:2024vxt}, although typically restricted to systems with a single spatial site~\cite{Klco:2018zqz,Illa:2022jqb}. While these studies validate the theoretical foundation for quantum simulations of field theories, scaling to larger volumes remains computationally expensive on gate-based platforms due to circuit depth, noise, and qubit overhead.

In this work, we present a framework for simulating scalar field theory on a quantum annealer, which broadens the scope of annealing-based importance sampling to include models with real-valued scalar fields. We construct field configurations $\phi(x)$ by mapping the theory into a format compatible with quantum annealing and apply the Metropolis-Hastings algorithm to perform sampling over large lattice volumes. A key technical obstacle is the quartic interaction terms in the field potential, which cannot be directly implemented on current quantum annealers as they require quadratic Hamiltonians.  To resolve this, we introduce a set of polynomial reduction techniques that introduce auxiliary qubits to systematically transform quartic potentials into effective quadratic Hamiltonians.

We evaluate three reduction techniques by measuring their fidelity and sampling efficiency, and benchmark the resulting simulations against classical Monte Carlo techniques. Our results demonstrate that, despite current hardware constraints, quantum annealers can achieve efficient sampling of scalar field configurations and offer a promising computational framework for scalable studies of quantum field theories.

This paper is organized as follows. Section~\ref{sec:QUBO} describes the construction of the quadratic unconstrained binary optimization (QUBO) matrix and introduces the polynomial reduction techniques and analyzes the distributions generated by the quantum annealer. In Sec.\ref{sec:results}, we present results from large-volume simulations performed using the Metropolis-Hastings algorithm with distributions obtained from the D-Wave quantum annealer, compare them with classical Metropolis Monte Carlo simulations, and discuss the scalability of the embedding strategies employed on the D-Wave platform.

\section{Variable digitization and polynomial reduction method\label{sec:QUBO}}
Consider the lattice model for the $\phi^4$ field theory in two-dimensional Euclidean space. Here, the scalar field $\phi(x)$ is defined at each point $x \in \Lambda$ on the lattice $\Lambda$. This lattice model is described by the action
\begin{align}
    \mathcal{S} = \sum_{x \in \Lambda} \bigg[& \frac{1}{2}\sum_\nu(\phi(x+a e_\nu) - \phi(x))^2 + \frac{1}{2}\mu_0^2\phi(x)^2 \nonumber \\&+ \frac{1}{4}\lambda_0\phi(x)^4  \bigg]\ ,
\end{align}
where $a$ is the lattice spacing and $e_\nu$ denotes the unit vector along $\nu=1,2$ direction. 
%
%
The lattice action and its parameters can also be rewritten as
\begin{align}\label{eq:phi4-action}
    S=&-2\kappa \sum_{\langle x, y \rangle}\phi(x)\phi(y) \nonumber \\&+ \sum_{x}\left(\phi(x)^2 + \lambda(\phi(x)^2-1)^2\right)\,,
\end{align}
where $\kappa$ is the hopping parameter and $\langle x, y \rangle$ denotes adjacent sites. 
In the regime of $\kappa\rightarrow 0$ with large $\lambda$,  the model lies in a $\mathbb{Z}_2$-symmetric phase with negligible coupling between adjacent lattice sites, which leads to a vanishing scalar expectation value $\langle\phi \rangle=0$. As $\kappa$ increases while keeping $\lambda$ fixed but finite, field interactions between the neighboring sites become strong and eventually reaches the critical point where the correlation length diverges.  Beyond this threshold, the system experiences a phase transition to a symmetry-broken phase. This behavior near criticality resembles that of the 2d Ising model near its critical temperature. We aim to implement the model on a quantum annealer and compare its critical behavior to known Ising universality.

Quantum annealers, which are currently accessible and programmable, offer a practical method for annealing an Ising-type Hamiltonian of qubits, as depicted by
\begin{align}
    H = \sum_{i} h_i \sigma_i^z + \sum_{i,j}J_{ij}\sigma_i^z\sigma_j^z \ ,
\end{align}
where $\sigma_i^z$ denotes the Pauli $z$-matrix for $i$-th spin. The parameters $h_i$ and $J_{ij}$ denote the spin bias and the coupling between two spins, respectively. These parameters are programmable and are utilized to encode the target quantum systems. 
The quantum processor within the annealing device is designed to find the ground state of this Ising spin system, thereby determining the lowest energy state of our objective quantum systems. 

Therefore, to realize the scalar field theory on a lattice in a quantum simulator, it is necessary to transform the lattice action in Eq.\eqref{eq:phi4-action} into the Ising Hamiltonian form, or equivalently, into the QUBO problem formulation:
\begin{align}
    E = \sum_i Q_{ii}q_i + \sum_{i<j}Q_{ij}q_iq_j \ ,
\end{align}
where $q_i$ are binary variables that take values of either $0$ or $1$ for qubits, and $Q_{ij}$ denotes an upper-triangular matrix.

We will now introduce QUBO formulations for the lattice model of the scalar field theory. We will first present a QUBO formulation utilizing the standard polynomial reduction method, enabling us to substitute the quartic interaction terms in the lattice action with quadratic terms. Next, we will propose an enhanced QUBO formulation that preserves the $\mathbb{Z}_2$ symmetry, which was explicitly broken in the standard polynomial reduction process. Finally, we will introduce a third method that uses extra qubit for sign in addition to the standard method.

\subsection{Method I: Standard method\label{sec:method1}}

In order to convert the lattice action in Eq.\eqref{eq:phi4-action} into a QUBO Hamiltonian, we first digitize the real scalar value $\phi(x)$ at each lattice site in terms of binary variables $q_i$. Assuming that we allocate $n_q$ bits for each scalar value, we can express the scalar field as
\begin{align}
    \phi(x) &= \phi_{min} + \delta \sum_{n=0}^{n_q-1} 2^n q_n(x)\,,  \\
    \delta&=\frac{\phi_{max} - \phi_{min}}{2^{n_{q}}-1} \ , \nonumber
\end{align}
with $n_q$ binary variables $q_n(x)$ at $x\in \Lambda$. The precision of the scalar field in this digitization process is determined by the values of $\phi_{min}, \phi_{max}$, and the number of qubits allocated per site $n_q$. By increasing $n_{q}$, we can enhance the precision but it also increases the QUBO matrix size, making it more challenging to achieve optimal solutions.
Adjustments of $\phi_{min}$ and $\phi_{max}$ can also improve the resolution, as demonstrated by the zooming technique discussed in \cite{Illa:2022jqb}.

The major challenge in converting to the QUBO Hamiltonian, which allows only up to quadratic interactions among binary variables, stems from the presence of quartic interactions of the scalar fields in the lattice action. For this, we employ the polynomial reduction method for binary variables. 
This approach involves utilizing the quadratic penalty function, as described in \cite{BOROS2002155}
\begin{align}\label{eq:penalty1}
    P(q_1,q_2;z) = q_1q_2 - 2(q_1+q_2)z + 3z \ .
\end{align}
This function reaches its minimum value, $P=0$, when $q_1q_2=z$ for two binary variables $q_1,q_2$ and an auxiliary binary variable $z$, whereas $P>0$ when $q_1q_2\neq z$. Table~\ref{tab:penalty} illustrates the values of the penalty function for given binary variables $q_1,q_2$ and $z$.
This means that adding the penalty function, with a suitable weight, to the QUBO Hamiltonian enables the quantum annealer to find optimal solutions minimizing $P$, which thus ensures the constraint $q_1q_2=z$. Therefore, by utilizing the penalty function, we can reduce the quadratic terms such as $q_1q_2$ into linear terms in terms of auxiliary variables like $z$.
\begin{table}[hpt]
     \centering
     \begin{tabular}{|c|c|| c | c|}
        \hline
        $q_1 \ q_2 \ z$ & $P(q_1,q_2;z)$ & $q_1 \ q_2 \ z$ & $P(q_1,q_2;z)$ \\
        \hline
        0 \ 0 \ 0 & 0 & 0 \ 0 \ 1 & 3 \\
        0 \ 1 \ 0 & 0 & 0 \ 1 \ 1 & 1 \\
        1 \ 0 \ 0 & 0 & 1 \ 0 \ 1 & 1 \\
        1 \ 1 \ 1 & 0 & 1 \ 1 \ 0 & 1 \\
        \hline
     \end{tabular}
     \caption{Penalty function $P(q_1,q_2;z)$ for binary variables $q_1,q_2$ and $z$.}
     \label{tab:penalty}
\end{table}

Practically, we substitute all quadratic terms of the form $q_m(x) q_n(x)$, which appear in the digitization of the quadratic term $\phi(x)^2$, with auxiliary variables $z_{mn}(x)$ and add the following penalty terms into the QUBO Hamiltonian:
\begin{align}
    E = S(q;z) + w P(q;z) \ ,
\end{align}
where $w$ is a positive weight. Then the lattice action that includes quartic interaction terms can be reformulated into the QUBO Hamiltonian format, and can be implemented on the quantum annealer. 

In our simulation, we have implemented our model on D-Wave's most advanced quantum annealer to date, the Advantage2 prototype. This device is built on the Pegasus topology that provides 15 to 20-way connectivity for each physical qubit. In the QUBO Hamiltonian for the lattice scalar theory, the $n_q(n_q+1)/2$ qubits representing a single scalar field at a lattice site requires all-to-all connectivity among them. However, the level of connectivity offered by the Pegasus topology is insufficient for a direct mapping of each logical qubit from the lattice scalar model to an individual physical qubit on the annealer. Therefore, to properly embed our model, we must map each logical qubit to multiple physical qubits.

\begin{figure*}[hpt]
    \subfigure[Validity rate versus penalty factor $w$\label{fig:asym_w}]{
    \includegraphics[width=0.45\linewidth]{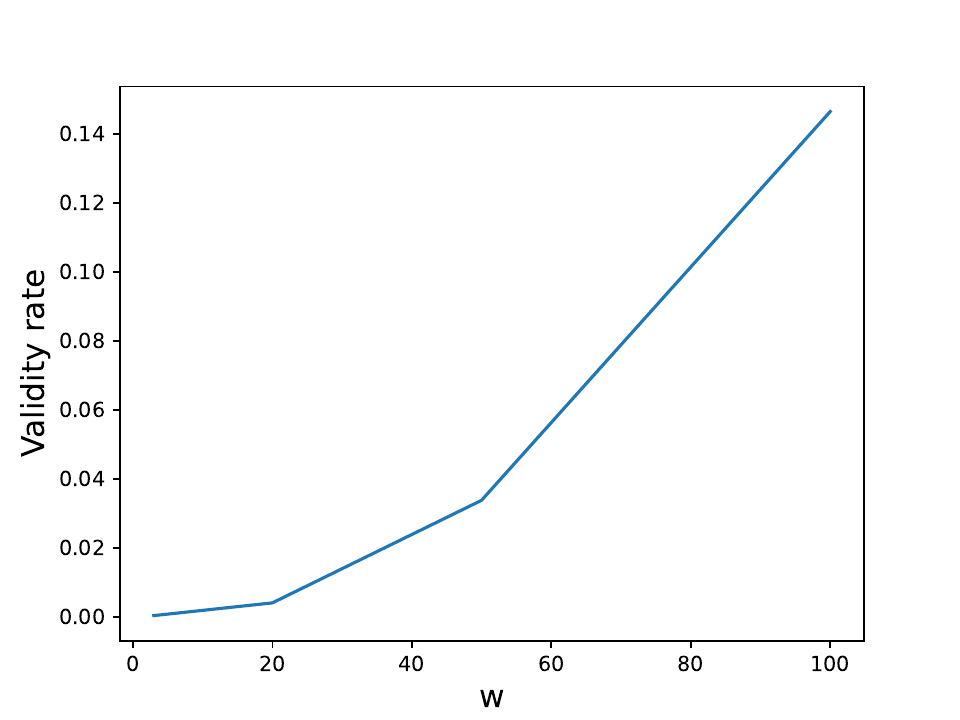}
    }
    \subfigure[Distribution of valid configurations at $\lambda=10$, $\kappa=0$, $n_{qubit}=6$ and $w=50$\label{fig:asym_dist}]{
    \includegraphics[width=0.45\linewidth]{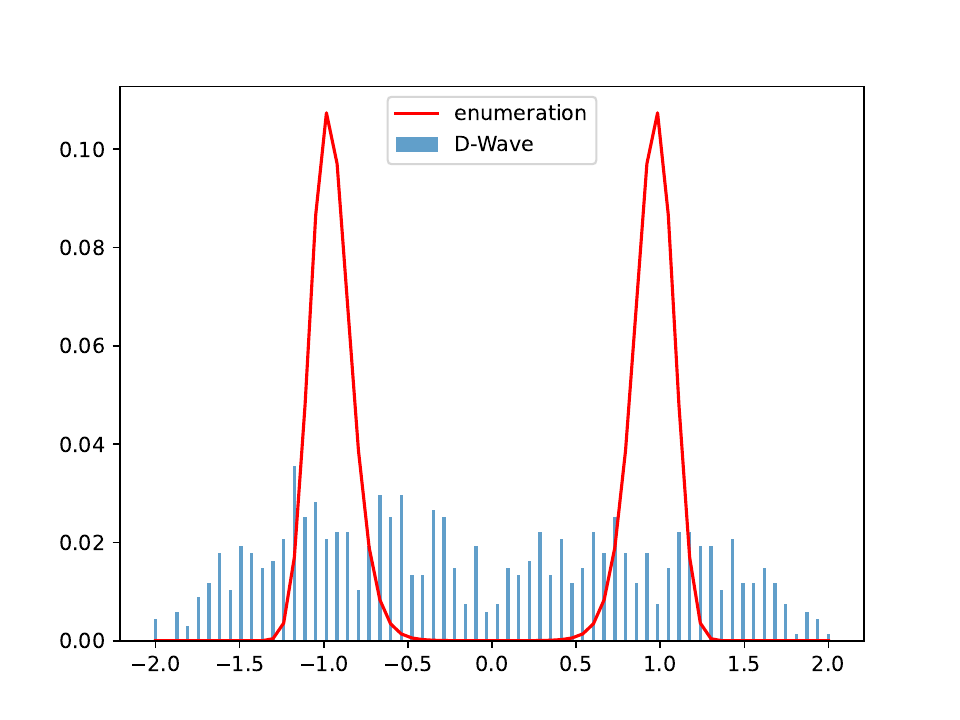}
    }
    \caption{(a): Validity rate as a function of the penalty factor $w$. As increasing $w$ the validity rate increases. (b): the distribution at $\lambda=10$, $\kappa=0$.  By increasing the penalty factor $w$, the action part of the QUBO becomes negligible. Hence, D-Wave can not generate the desired distribution. The number of qubits is 6. The enumeration is computed using the same digitized $\phi$ variable.}
\end{figure*}

We first investigate the dependence of the validity rate on the penalty factor in order to maximize the number of valid configurations that satisfy the constraint given in Eq.\eqref{eq:penalty1}. The chain strength, another important optimization parameter, is fixed at its optimal value of $0.4$; its dependence will be discussed separately in Sec.\ref{sec:mehod3}. The validity rate is defined as the ratio of solution vectors that satisfy the constraint to the total number of solution vectors measured. As shown in Fig.\ref{fig:asym_w}, the validity rate increases with the penalty factor $w$. This behavior reflects the role of the penalty term in balancing the constraint against the action: a large $w$ enforces the constraint more strongly, yielding a higher validity rate, but suppresses the contribution of the action. Conversely, a small $w$ is necessary for importance sampling but results in fewer valid solutions. Thus, choosing an appropriate value for $w$ requires balancing these competing effects and remains a nontrivial task in this setting.

As shown in Fig.\ref{fig:asym_dist}, the distribution of valid solutions generated by D-Wave completely fails to capture the distribution computed by enumeration (red line). For lower values of $w$, due to the extremely low validity rate, it was not possible to accumulate a sufficient amount of data to build a distribution of valid configurations. This failure is largely due to our standard digitization method, which explicitly breaks the $\mathbb{Z}_2$ symmetry of the lattice action.

In the next sections, we modify the digitization method and structure of the penalty factor and obtain the reasonable validity rate to construct the histogram for the simulations.

\subsection{Method II: Enhanced penalty factor method\label{sec:method2}}
In this section, we present an alternative embedding approach for the lattice scalar model into the QUBO Hamiltonian with enhanced penalty factor. This method uses two auxiliary qubits to convert the quartic to quadratic, but the validity rate is large enough at small penalty factor, so the weight matrix also plays a role in generating the distribution for given potential. 

We propose the following penalty function:
\begin{align}
    \tilde{P}(q_1,q_2;z,s) =& 1 -q_1-q_2+2q_1q_2 -2(q_1+q_2)z \nonumber \\ &+3z+4(q_1+q_2-z)s \ ,
\end{align}
with two auxiliary variables $z$ and $s$. The respective values of this penalty function for given binary variables are listed in Table~\ref{tab:penalty2}. Notably, when the penalty function is minimized, i.e. when $\tilde{P}=0$, the binary variables satisfy
\begin{align}
    q_1+q_2+z = 1 \ {\rm mod} \ 2 \ .
\end{align}
\begin{table*}[h]
     \centering
     \begin{tabular}{|c|c|| c | c|}
        \hline
        $q_1 \ q_2 \ z \ s $ & $\tilde{P}(q_1,q_2;z,s)$ & $q_1 \ q_2 \ z \ s$ & $\tilde{P}(q_1,q_2;z,s)$ \\
        \hline
        0 \ 0 \ 0 \ 0 & 1 & 0 \ 0 \ 1 \ 0& 4 \\
        0 \ 1 \ 0 \ 0 & 0 & 0 \ 1 \ 1 \ 0& 1 \\
        1 \ 0 \ 0 \ 0 & 0 & 1 \ 0 \ 1 \ 0& 1 \\
        1 \ 1 \ 0 \ 0 & 1 & 1 \ 1 \ 1 \ 0 & 0 \\
        0 \ 0 \ 0 \ 1 & 1 & 0 \ 0 \ 1 \ 1  & 0 \\
        0 \ 1 \ 0 \ 1 & 4 & 0 \ 1 \ 1 \ 1  & 1 \\
        1 \ 0 \ 0 \ 1 & 4 & 1 \ 0 \ 1 \ 1  & 1 \\
        1 \ 1 \ 0 \ 1 & 9 & 1 \ 1 \ 1 \ 1  & 4 \\
        \hline
     \end{tabular}
     \caption{New penalty function $\tilde{P}(q_1,q_2;z,s)$ for binary variables $q_1,q_2$ and auxiliary variables $z,s$.}
     \label{tab:penalty2}
\end{table*}
This suggests that, for solutions with $\tilde{P}=0$, we can replace the product of two binary variables with a linear sum of binary variables, as follows
\begin{align}\label{eq:replace2}
    q_1q_2 \  \rightarrow \ \frac{q_1+q_2+z-1}2 \ .
\end{align}
This provides an alternative polynomial reduction method. One notable aspect of this method is that the replacement rule in Eq.\eqref{eq:replace2} aligns with the $\mathbb{Z}_2$ action, which simultaneously flips $q_1$ and $q_2$, at $\tilde{P}=0$. 
Hence, by employing the alternative polynomial reduction method along with the additional penalty term given by
\begin{align}
    \Delta E = w \sum_{x}\sum_{m<n}\tilde{P}(q_m(x),q_n(x);z_{mn}(x),s_{mn}(x)) \ ,
\end{align}
we can achieve the $\mathbb{Z}_2$ symmetric embedding of the lattice scalar model into the QUBO Hamiltonian. We will demonstrate in the next section that the outcomes from this embedding method after the annealing process exhibit the $\mathbb{Z}_2$ symmetry.

\begin{figure*}[h]
    \subfigure[Validity rate versus penalty factor $w$\label{fig:sym_valid}]{
    \includegraphics[width=0.45\linewidth]{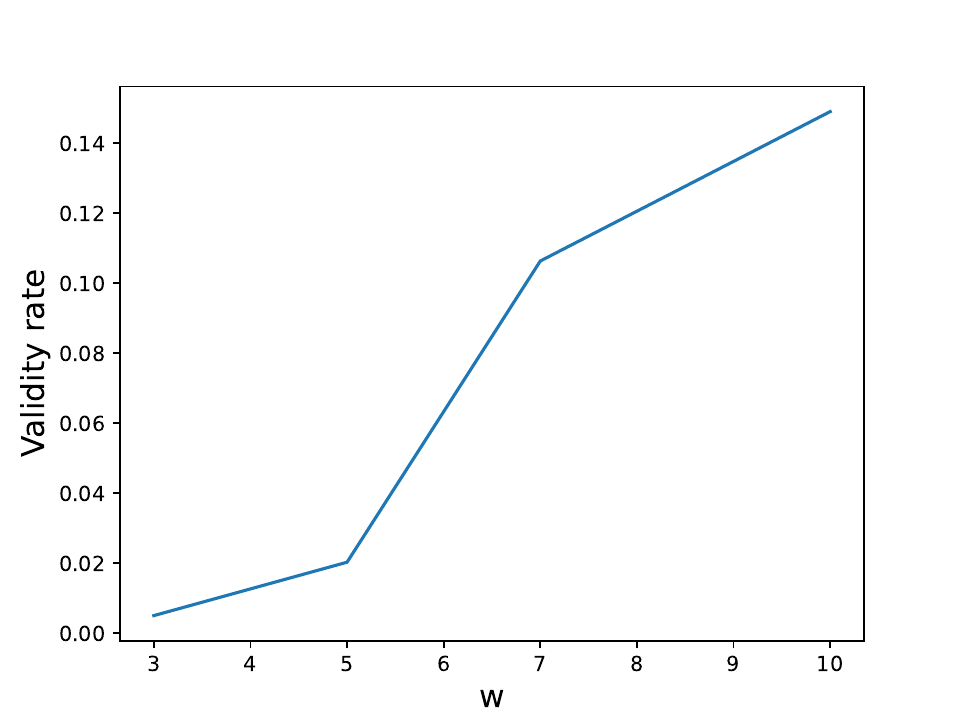}
    }
    \subfigure[Distribution at $\lambda=10$, KL divergence=1.2884\label{fig:sym_histo}]{
    \includegraphics[width=0.45\linewidth]{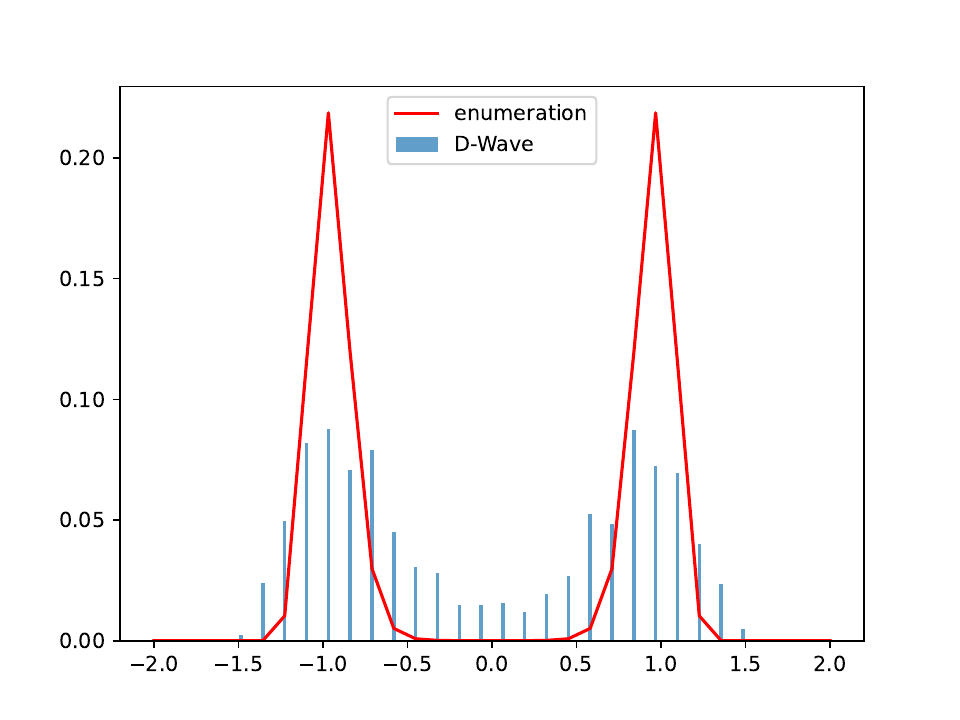}
    }
    \caption{(a): Validity rate as a function of the penalty factor $w$. As increasing $w$ the validity rate increases. (b): the distribution at $\lambda=10$, $\kappa=0$ and $w=10$. The number of qubits is 5, $\phi \in \{-2,2\}$. Although the Kullback-Leibler(KL) divergence of 1.2884 indicates non-negligible divergence, the overall shapes of the distributions remain comparable. The enumeration is computed using the same digitized $\phi$ variable.}
    
\end{figure*}
The validity rate at smaller values of $w$ is sufficient to accumulate meaningful statistics, as shown in Fig.\ref{fig:sym_valid}. We observe that the resulting distribution exhibits behavior consistent with the analytic prediction in Fig.\ref{fig:sym_histo}. For larger values of $w$, the KL divergence increases, and the distribution approaches a uniform distribution. We perform simulations on a $16 \times 16$ lattice using the Metropolis-Hastings algorithm with this distribution. In the ideal case, where the sampled distribution perfectly matches the analytic one, the acceptance rate would be unity. Therefore, we used the acceptance ratio as an indicator of the closeness between the sampled and analytic distributions. 
\begin{figure*}[h]
    \centering
    \subfigure[$w$ dependence of the acceptance rate \label{fig:acceptance_sym}]{
    \includegraphics[width=0.45\linewidth]{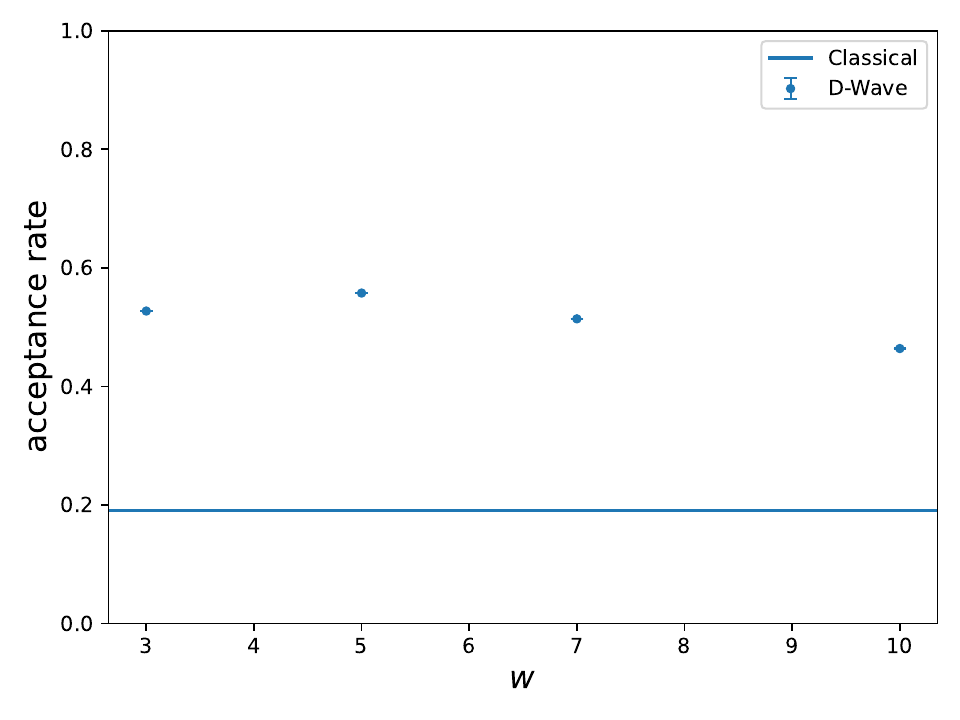}
    }
    \subfigure[$|\phi|$ at $\lambda=10.0$, $\kappa=0$ and comparison with classical Metropolis\label{fig:sym_absphi}]{
    \includegraphics[width=0.45\linewidth]{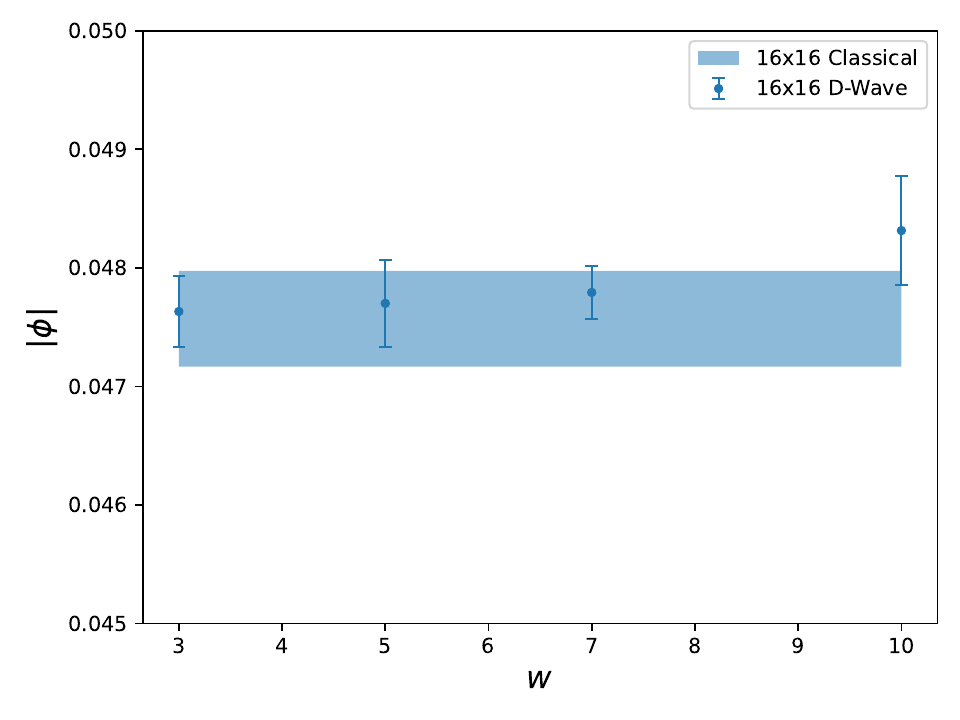}
    }
    \caption{(a): Acceptance rate versus penalty factor $w$ with 5 qubits and compare with classical Metropolis for $\lambda=10.0$ and $\kappa=0.0$, (b): the comparison of the results between classical Metropolis simulation and Metropolis-Hastings method with D-Wave.}
\end{figure*}

In Fig.\ref{fig:acceptance_sym}, we compute the acceptance rate of the Metropolis-Hastings algorithm using the D-Wave quantum annealer and compare it with that of the classical Metropolis method at $\lambda = 10$ and $\kappa = 0.0$. For the classical Metropolis algorithm, continuous values of $\phi$ were sampled from a uniform distribution over the interval $[-2,2]$ and the acceptance rate is approximately 20\%, whereas the Metropolis-Hastings method assisted by D-Wave achieves an acceptance rate of about 50\%. Furthermore, the expectation value of $|\phi|$ obtained from the quantum-assisted method is consistent with the results of the classical approach (Fig.\ref{fig:sym_absphi}). The details of the Metropolis-Hastings methods are described in Sec.\ref{sec:results}.

Each polynomial reduction method offers specific benefits and drawbacks. The $\mathbb{Z}_2$-symmetric approach presented in this subsection requires $n_q^2$ logical qubits per scalar field at a single site, including the auxiliary variables $z$ and $s$. This results in a significantly larger qubit overhead compared to the $n_q(n_q+1)/2$ logical qubits needed in the standard polynomial reduction method discussed earlier. However, as noted in the previous section, that earlier method suffers from a low validity rate, which prevents the generation of a reliable distribution.

In contrast, the $\mathbb{Z}_2$-symmetric method produces a substantially improved distribution. Compared with the classical Metropolis algorithm, it provides more efficient simulations while maintaining consistency with classical results.
In the next section, we propose a third method that preserves $\mathbb{Z}_2$ symmetry while requiring a comparable number of qubits to the standard polynomial reduction method.

\subsection{Method III: Standard method with extra sign qubit\label{sec:mehod3}}
We now propose a third digitization method that uses an extra sign qubit:
\begin{align}
    \phi(x)=(1-2 s) \times \bigg(\phi_{min} + \delta \sum^{n_q-1}_{n=0} 2^n q_q(x) \bigg)\,,
\end{align}
where $s$ controls the overall sign of $\phi$.
This embedding allows $\phi_{min} \ge 0$, which makes it possible to zoom in on narrower field ranges. 
The polynomial reduction and penalty factor function is the same as in Sec.\ref{sec:method1}.

\begin{figure*}[h]
    \centering
    \subfigure[Validity and acceptance rates vs \texttt{chain\_strength}\label{fig:V_chain_strength}]{
    \includegraphics[width=0.48\linewidth]{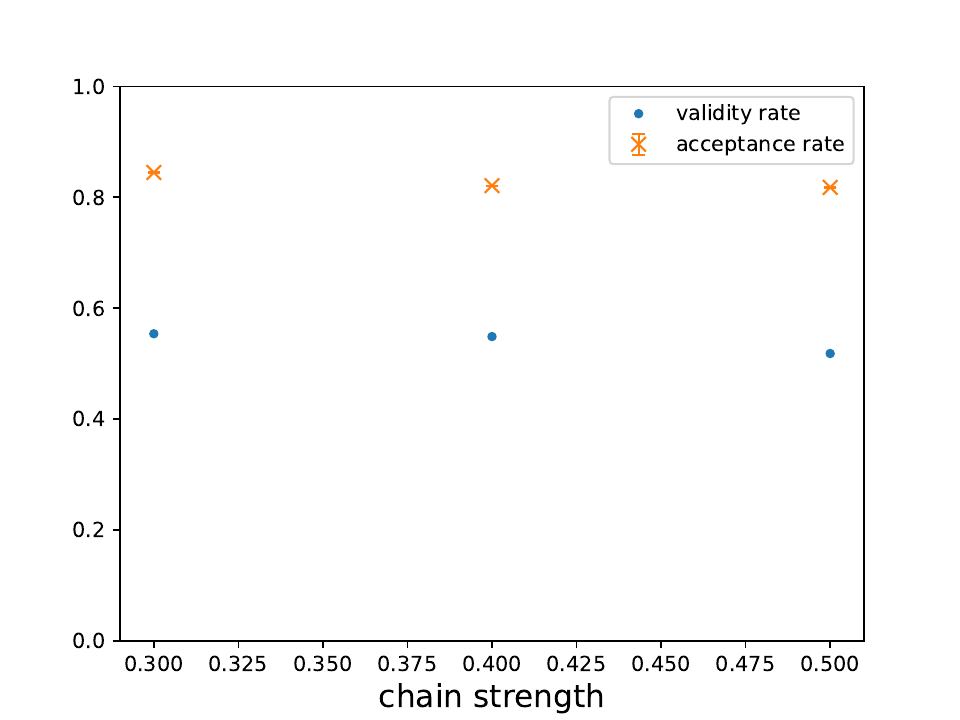}
    }
    \subfigure[Validity and acceptance rates vs penalty factor $w$\label{fig:V_w_sign}]{
    \includegraphics[width=0.48\linewidth]{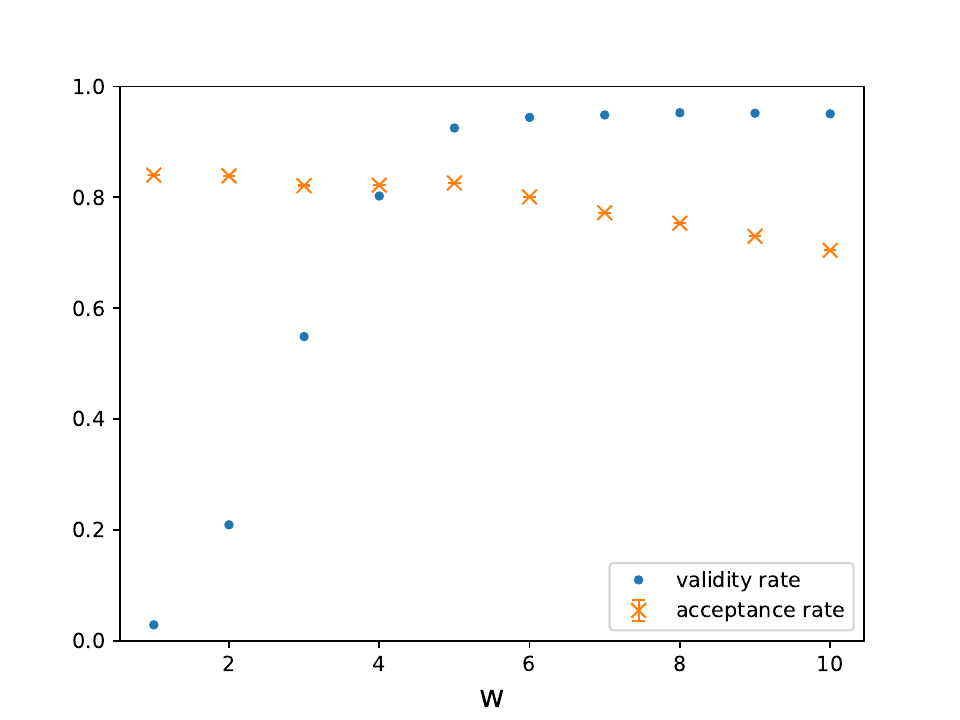}
    }
    \caption{\texttt{chain\_strength} and penalty factor $w$ dependence of validity and acceptance rates.}
    \label{fig:acceptance}
\end{figure*}

We optimize both the \texttt{chain\_strength} and the penalty factor $w$ to maximize the validity and acceptance rates, as shown in Fig.\ref{fig:acceptance}. 
The acceptance rate is computed using the Metropolis-Hastings algorithm. 
In this case, the optimal value of $w$ is significantly smaller than that required in the previous methods. 
Despite the reduced penalty factor, this method yields a substantially higher validity rate and a more accurate distribution, resulting in an acceptance rate of approximately 80\%.

\begin{figure*}[h]
    \centering
    \subfigure[At $\lambda=0.1$, KL divergence=0.0625]{
    \includegraphics[width=0.45\linewidth]{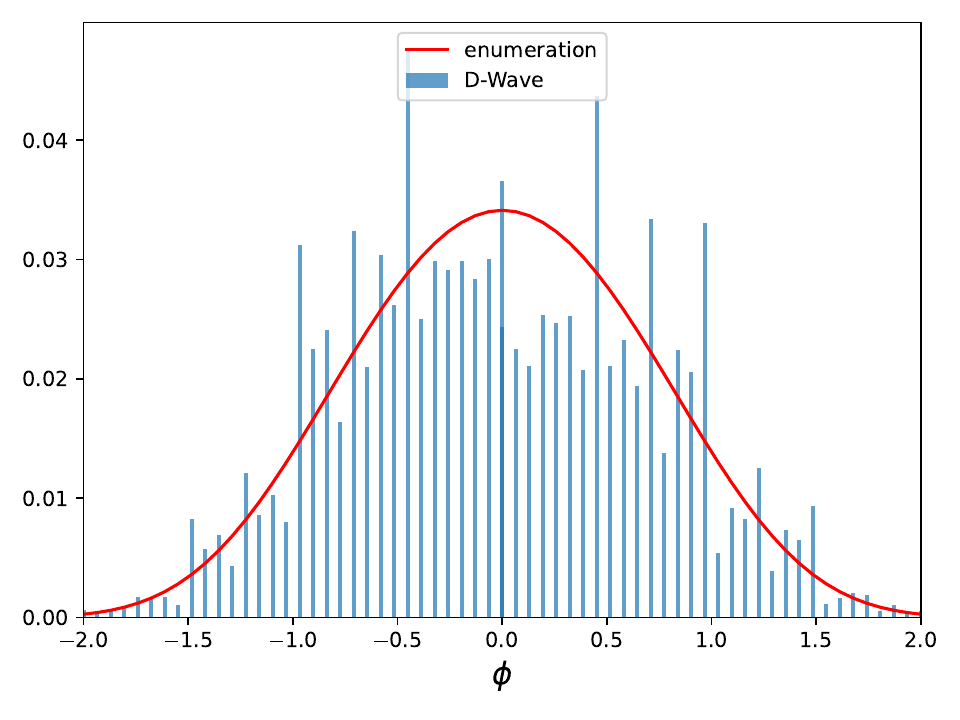}
    }
    \subfigure[At $\lambda=10$, KL divergence=0.0680]{
    \includegraphics[width=0.45\linewidth]{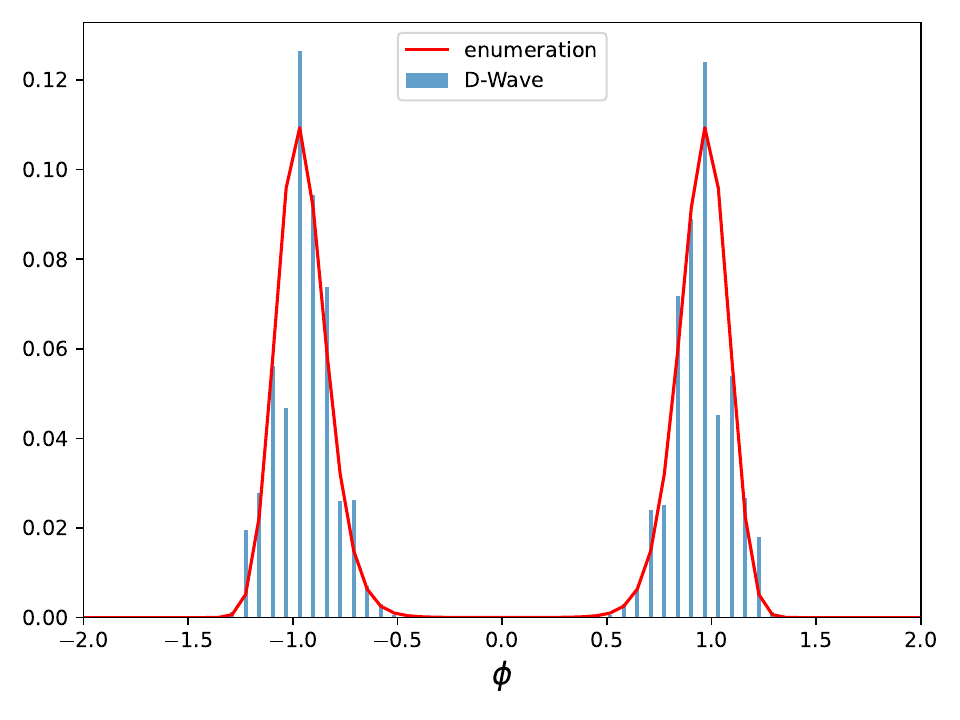}}
    \caption{The histogram of $\phi$ for $\lambda=0.1$ and $10$ with ($\kappa=0.0$, $n_{qubit}=5$ and one sign qubit, $w=3$, $\phi_{min}=0$, $\phi_{max}=2$) are shown. The red lines are the exact enumeration results. The KL divergences of 0.0625 and 0.0680 indicate a close agreement between the sampled and exact distributions.}
    \label{fig:histo}
\end{figure*}
To evaluate the fidelity of the quantum annealer-generated distribution Q (D-Wave) relative to the target Boltzmann distribution P (enumeration), we computed the Kullback-Leibler (KL) divergence. The resulting values, 0.0625 and 0.0680 for $\lambda=0.1, 10$ at $\kappa=0$, indicate that the two distributions are in close agreement. Since KL divergence quantifies the information loss when Q is used to approximate P, the small value observed here suggests that the annealer samples accurately reflect the desired statistical structure of the scalar field theory.

\section{Results\label{sec:results}}
In the absence of nearest-neighbor interactions ($\kappa = 0$), the lattice sites decouple and the field distribution reduces to that governed solely by the local potential. In this regime, we generate single site field distributions for various values of $\lambda$ and use them to sample lattice configurations via the Metropolis-Hastings algorithm. The acceptance probability for a proposed update from an old configuration to a new one is given by
\begin{align}
P = e^{-S_{\mathrm{new}} + S_{\mathrm{old}}} \frac{h_{\mathrm{old}}}{h_{\mathrm{new}}} ,
\end{align}
where $h_{\mathrm{old}}$ and $h_{\mathrm{new}}$ denote the heights of the respective field values in the precomputed histogram. If the histogram perfectly reproduces the Boltzmann distribution $e^{-S}$, the acceptance probability becomes unity, equivalent to a heatbath update. Thus, the acceptance rate provides a quantitative measure of how closely the sampling distribution approximates the ideal one.

\begin{figure*}[h]
    \centering
    \subfigure[Acceptance rate \label{fig:accept_lambda}]{
    \includegraphics[width=0.48\linewidth]{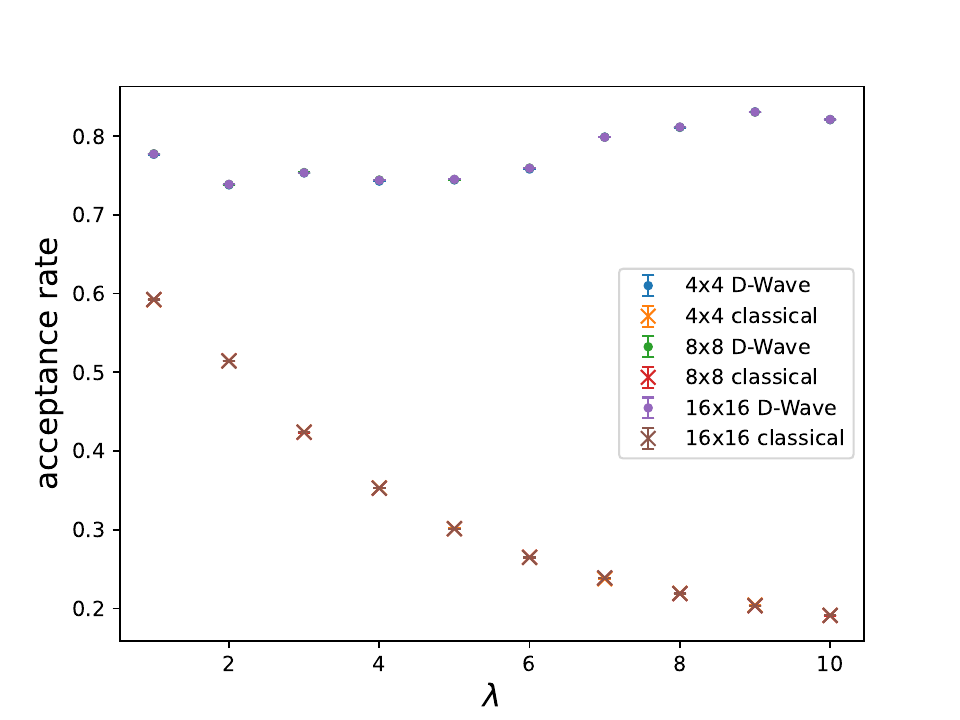}
    }
    \subfigure[$|\phi|$ versus $\lambda$ at $\kappa=0$ \label{fig:result_lambda}]{
    \includegraphics[width=0.48\linewidth]{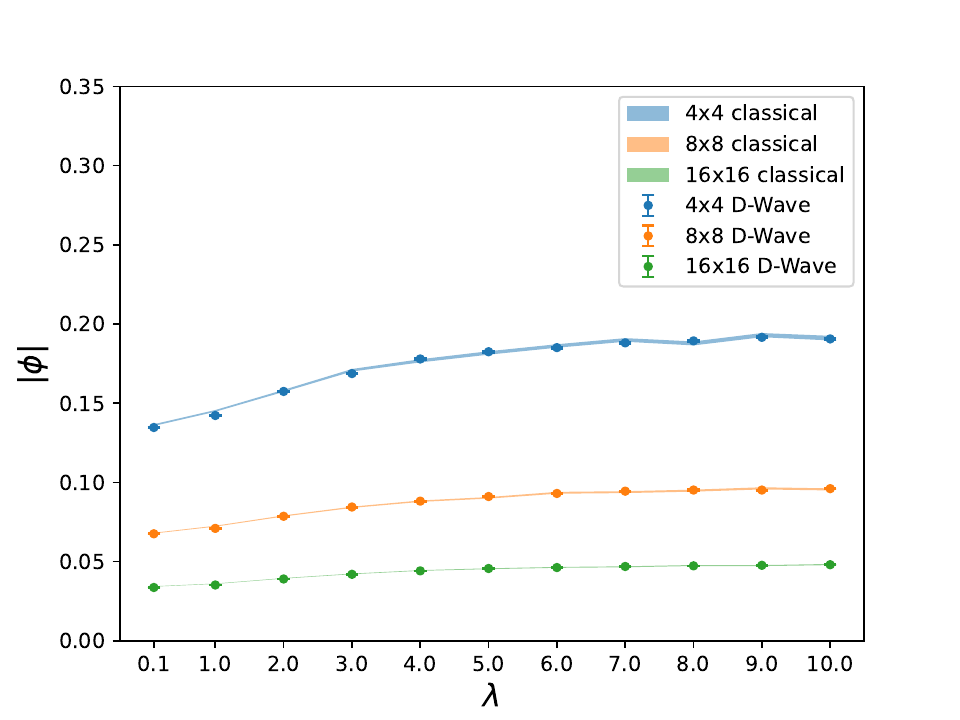}
    }
    \caption{(a) The acceptance rate versus $\lambda$ at $\kappa=0$ on $4 \times 4$, $8 \times 8$ and $16 \times 16$. Since all data points overlap, and there is no volume dependence. (b) The comparison of expectation value of $|\phi|$ between the classical simulation(bands) and D-Wave(data points).}
    \label{fig:lambda_dependence}
\end{figure*}

Figure~\ref{fig:accept_lambda} shows the acceptance rates for various lattice sizes, comparing the Metropolis-Hastings algorithm using distributions generated by the D-Wave quantum annealer with the standard classical Metropolis update. The acceptance rate exhibits a clear dependence on the coupling $\lambda$, with the classical Metropolis method performing poorly at large $\lambda$. In contrast, the D-Wave enhanced Metropolis-Hastings approach maintains an acceptance rate exceeding 70\% across the parameter range. Importantly, as shown in Fig.\ref{fig:result_lambda}, the expectation values of the observable $|\phi|$ obtained using both methods agree within statistical uncertainties.

\begin{figure*}
    \centering
    \subfigure[Acceptance rate \label{fig:accept_kappa}]{
        \includegraphics[width=0.48\linewidth]{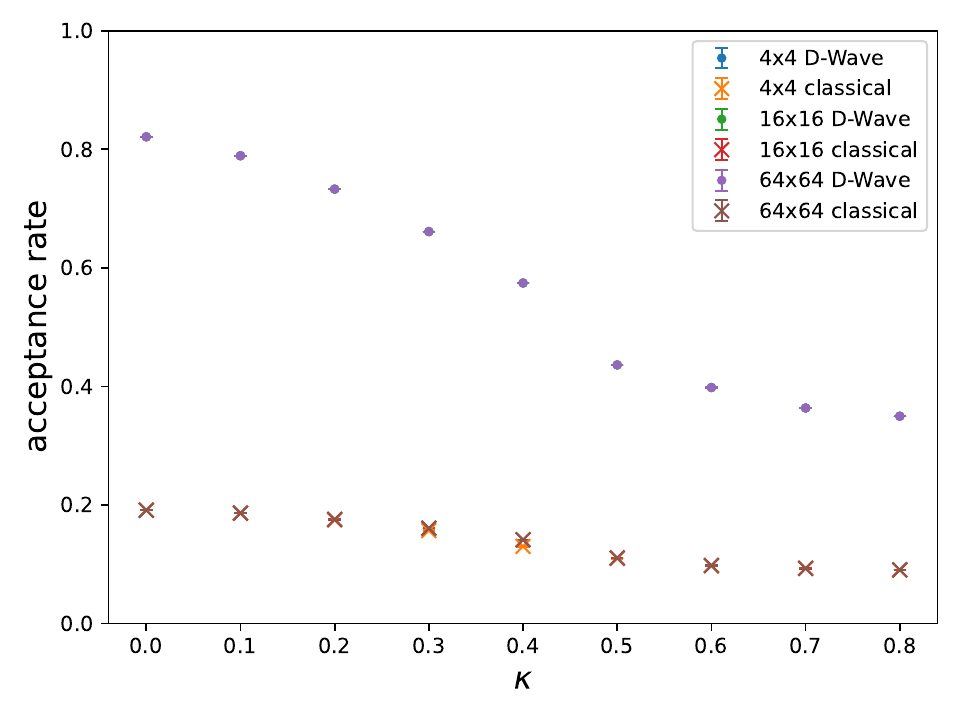}
    }
    \subfigure[$|\phi|$ versus $\kappa$ at $\lambda=10$ \label{fig:result_kappa}]{
        \includegraphics[width=0.48\linewidth]{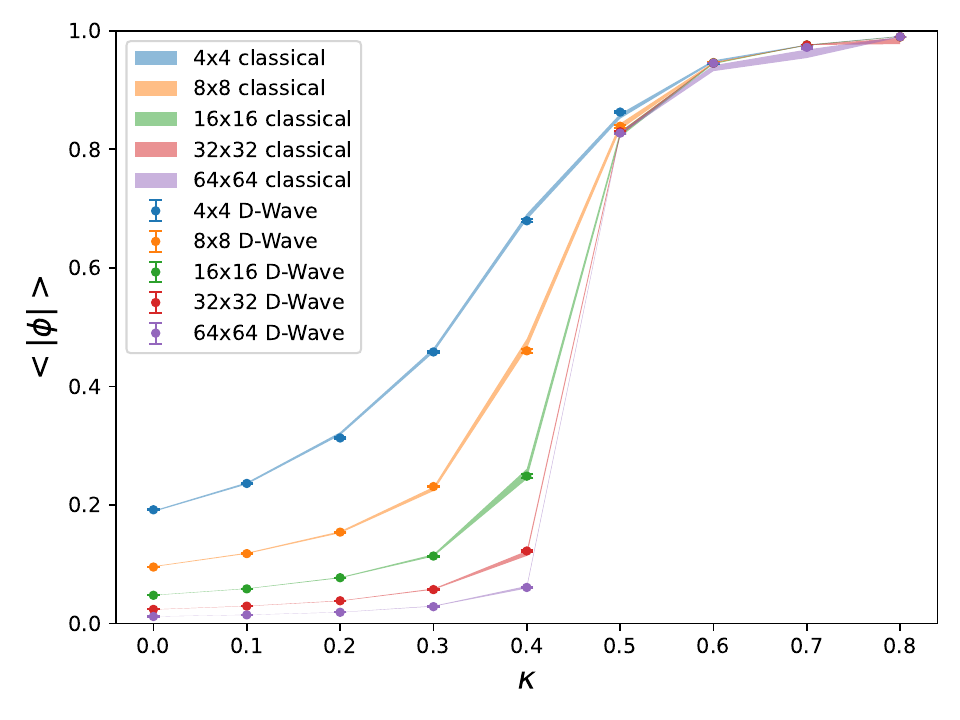}
    }
    \caption{(a) The acceptance rate versus $\kappa$ at $\lambda=10$ on $4 \times 4$, $16 \times 16$ and $64 \times 64$. No volume dependence is found. (b) The comparison of expectation value of $|\phi|$ between classical simulations and the method III using D-Wave on $4 \times 4$, $8 \times 8$, $16 \times 16$, $32 \times 32$ and $64 \times 64$ lattices.}
\end{figure*}

While the distribution was initially generated at $\kappa = 0$, it can still be utilized in simulations with finite $\kappa$. Nevertheless, due to the discrepancy between the generated and the target distributions at nonzero $\kappa$, the acceptance rate deteriorates as $\kappa$ moves away from zero as shown in Fig.\ref{fig:accept_kappa}.

We present the simulation results on $4 \times 4$, $8 \times 8$, $16 \times 16$, $32 \times 32$ and $64 \times 64$ for both D-Wave and classical Metropolis method in Fig.\ref{fig:result_kappa}. A phase transition to a symmetry broken phase is observed near the critical point $\kappa_c$. The behavior of this model near the critical point indeed resembles that of the 2D Ising model near its transition temperature. 

\begin{figure*}
    \centering
    \subfigure[Susceptibility of $|\phi|$\label{fig:sus}]{
    \includegraphics[width=0.45\linewidth]{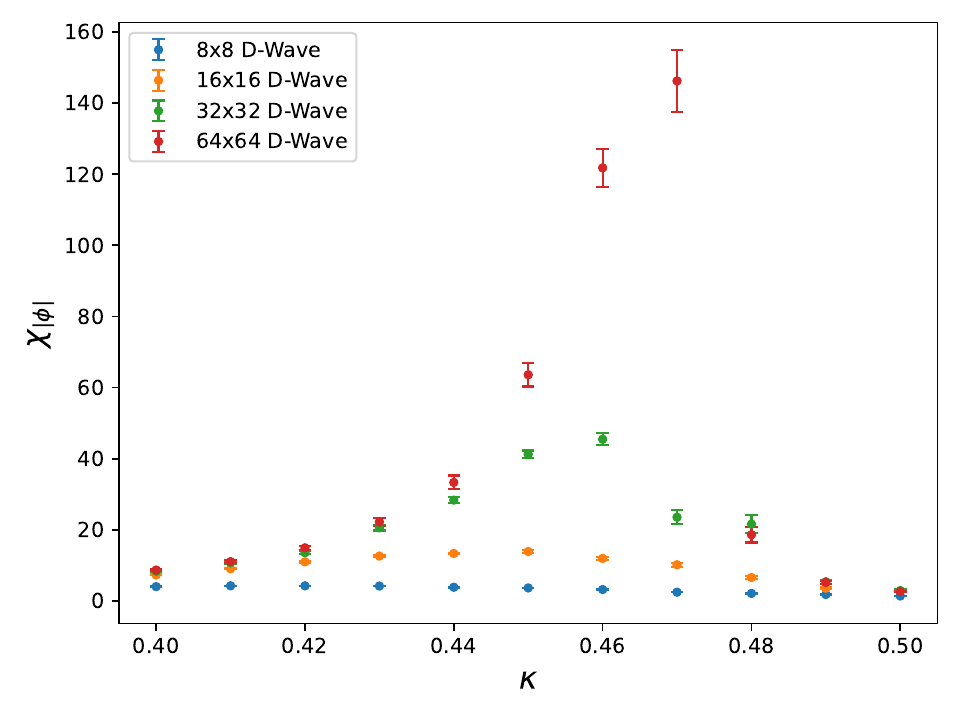}
    }
    \subfigure[Skewness $B_3^{|\phi|}$ \label{fig:B3}]{
    \includegraphics[width=0.45\linewidth]{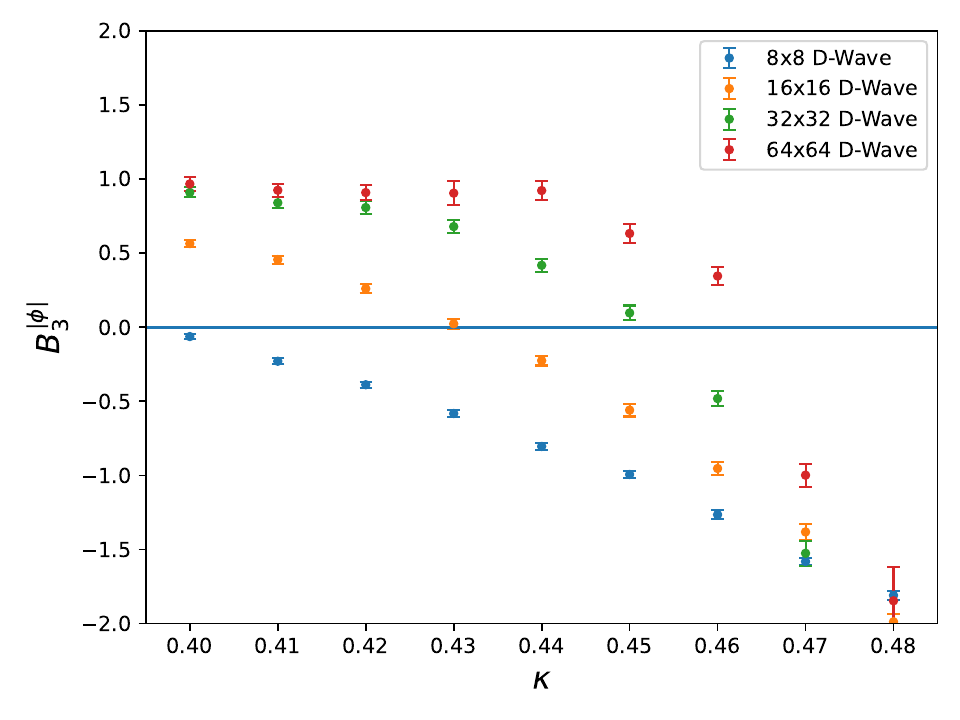}
    }
    \caption{Susceptibility and skewness $B_3$ of the observable $|\phi|$ at $\lambda=10$ obtained by method III.}
\end{figure*}

We have computed both the susceptibility $V(\langle \phi(x)^2 \rangle - \langle \phi(x) \rangle^2)$ (Fig.\ref{fig:sus}) and the $B_3$ defined by
\begin{align}
    B_3 = \frac{\langle (\delta O )^3\rangle}{\langle (\delta O )^2\rangle^{3/2}} = \frac{\langle O^3 \rangle - 3\langle O^2 \rangle\langle O \rangle +2\langle O \rangle^3}{(\langle O^2 \rangle-\langle O \rangle^2)^{3/2}}
\end{align}
as shown in Fig.\ref{fig:B3}, using Method III. The pseudo critical point $\kappa_c\sim 0.47$ is determined as the point where the skewness crosses zero. The divergence of the susceptibility at this critical coupling (shown in Fig.~\ref{fig:sus}) supports this determination. 

\begin{figure*}
    \centering
    \subfigure[$\kappa < \kappa_c$]{
    \includegraphics[width=0.45\linewidth]{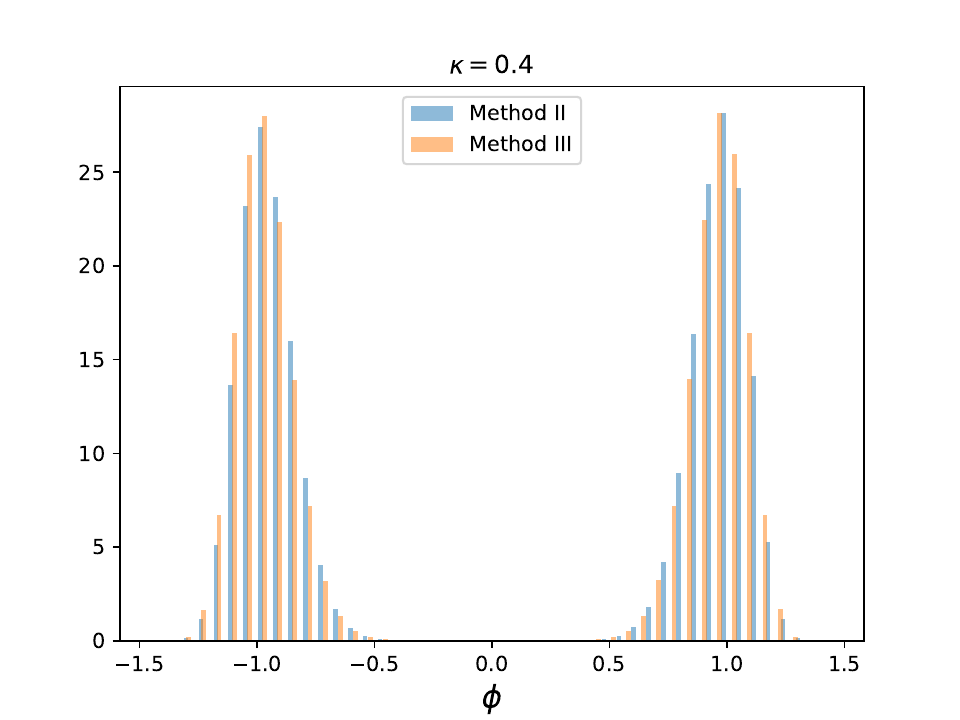}
    }
    \subfigure[$\kappa > \kappa_c$]{
    \includegraphics[width=0.45\linewidth]{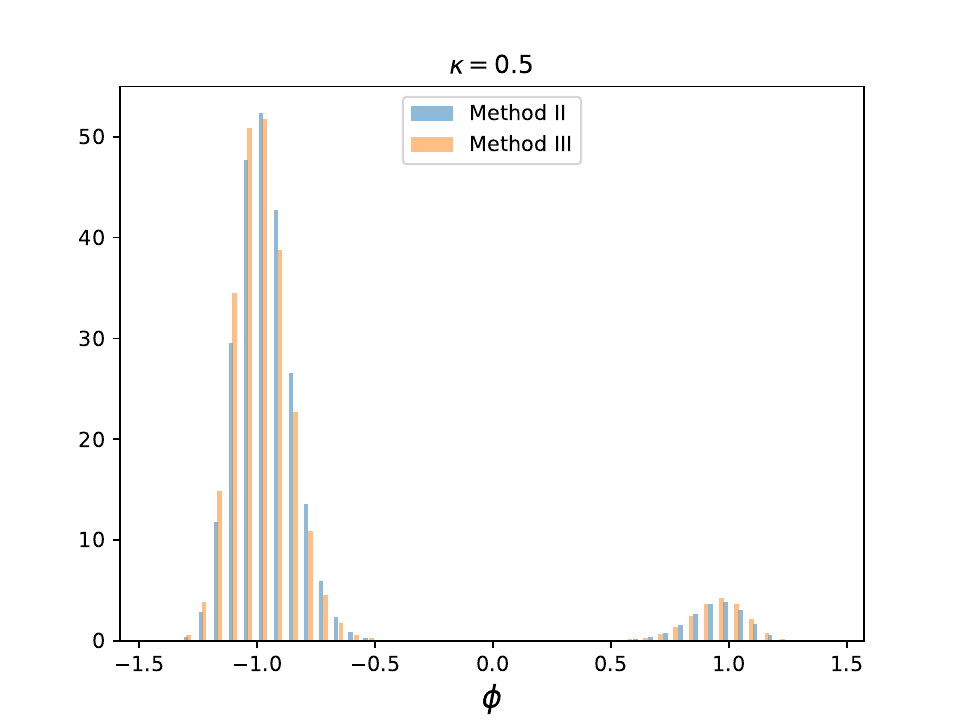}
    }
    \caption{Histograms of observable $\phi$ at $\lambda=10$, $\kappa=0.4,0.5$. The first reflects the $Z_2$ symmetric phase; the second shows that the symmetry is broken.}
    \label{fig:histo_kappa_c}
\end{figure*}

Scalar field distributions obtained from Methods II and III at couplings below and above $\kappa_c$ are compared in Fig.~\ref{fig:histo_kappa_c}. Both methods consistently produce symmetric peaks at $\phi=\pm1$ when $\kappa < \kappa_c$, indicative of the disordered phase. Conversely, at couplings beyond $\kappa_c$, the distributions clearly signal the breaking of $\mathbb{Z}_2$ symmetry, which is the characteristic of the ordered phase. Here, the role of the magnetization in the Ising model is effectively played by the observable $\langle|\phi|\rangle$.

To map out the phase structure of the two dimensional lattice scalar field theory, we numerically determined the critical line $\kappa_c(\lambda)$ separating the symmetric and spontaneously broken phases for a range of quartic couplings, $\lambda \in [0.01, 10]$. The bare potential, given by
$V(\phi) = \phi^2 + \lambda(\phi^2 - 1)^2$, exhibits qualitatively distinct features depending on the value of $\lambda$, which in turn directly influence the behavior of $\kappa_c(\lambda)$.

In the very weak coupling regime ($\lambda \ll 1$), the quartic term becomes negligible, and the potential is effectively dominated by the quadratic term. In this limit, the restoring force near the origin is weak, and fluctuations are less confined to $\phi = 0$, allowing field configurations to explore larger amplitudes. As a result, spontaneous symmetry breaking can occur even at very small values of $\kappa$. As $\lambda$ increases, the quartic interaction gradually strengthens, providing additional stabilization near the origin and suppressing large fluctuations. Around $0.01 \lesssim \lambda \lesssim 0.2$, the potential becomes nearly flat near $\phi = 0$, leading to enhanced fluctuations and a moderate increase in $\kappa_c(\lambda)$, since a stronger kinetic term is required to destabilize the symmetric phase.

For intermediate couplings ($0.1 \lesssim \lambda \lesssim 1$), the potential remains nearly flat but does not yet develop a clear double well structure. In this region, $\kappa_c(\lambda)$ forms a broad plateau, reflecting the insensitivity of the critical behavior to small changes in $\lambda$. The balance between kinetic and potential terms stabilizes the phase boundary across this range despite large field fluctuations.

At strong coupling ($\lambda \gtrsim 1$), the potential develops a pronounced double well shape with local minima at $\phi = \pm 1$. This signals the onset of the symmetry breaking phase dominated by the quartic interaction. As the wells deepen, the field becomes increasingly localized near the minima, and spontaneous symmetry breaking occurs more readily. Accordingly, the critical hopping parameter $\kappa_c(\lambda)$ decreases with increasing $\lambda$, eventually saturating as the system approaches the Ising limit.

\begin{figure}
    \centering
    \includegraphics[width=\linewidth]{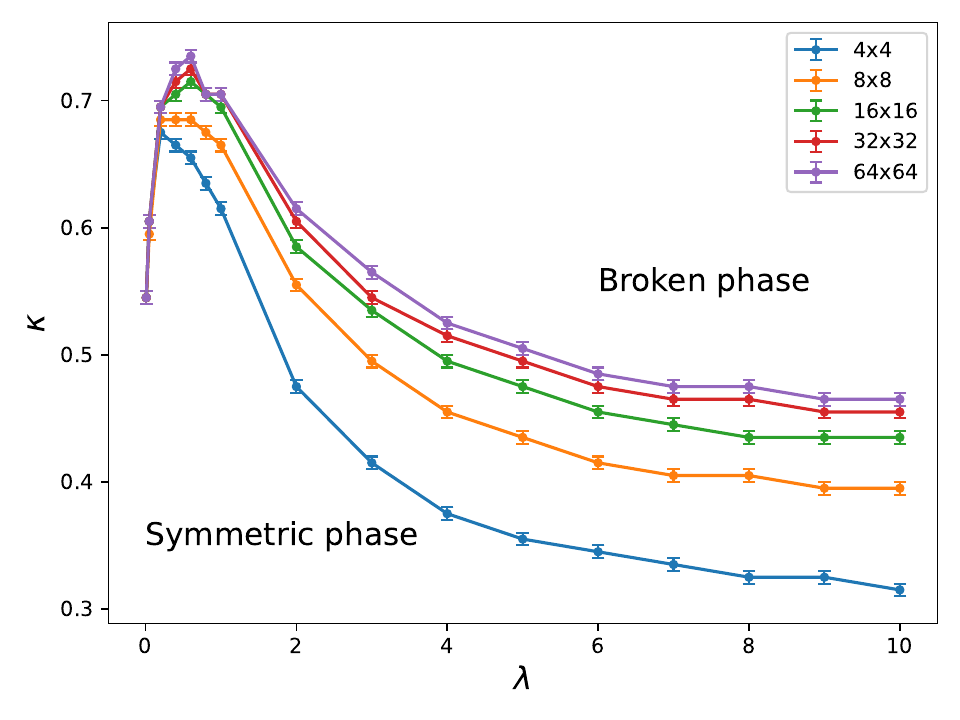}
    \caption{Phase boundary in the $\lambda-\kappa$ plane. The critical line $\kappa_c(\lambda)$ separates the symmetric phase from the spontaneously broken phase. The pseudo critical points $\kappa_c$ are determined from $B_3$ by scanning $\kappa$ with a precision of $0.01$. We used $\phi_{max}=10$ for $\lambda < 2$, and $\phi_{max}=2$ otherwise.}
    \label{fig:phase_boundary}
\end{figure}

Figure~\ref{fig:phase_boundary} shows the resulting phase boundary in the $(\lambda, \kappa)$ plane. The observed nonmonotonic behavior of $\kappa_c(\lambda)$, which initially increases, then plateaus, and finally decreases, captures the crossover between qualitatively distinct regimes of the bare potential and provides a clear signature of the underlying changes in field dynamics. The shape of the phase boundary thus smoothly interpolates between the Gaussian fixed point and the Ising-like behavior at large $\lambda$. The extracted values of $\kappa_c(\lambda)$ offer a quantitative characterization of this crossover and serve as a useful benchmark for studies of critical phenomena and renormalization group flows in low-dimensional scalar field theories.

We compare the autocorrelation times obtained from classical simulations and the D-Wave method in Table~\ref{tab:autocor}. Simulation details and statistics are provided in Appendix~\ref{app:stats}. For this comparison, we select $\kappa = 0.5$\footnote{The choice of $\kappa = 0.5$ is motivated by the fact that at $\kappa_c$, the correlation time diverges, making fair comparisons difficult with finite statistics. $\kappa=0.5$ is close to $\kappa_c$, which is an interesting region to study while still allowing for meaningful statistical analysis.}, which lies near the critical point at $\lambda = 10$ and above the critical value $\kappa_c$ where the $Z_2$ symmetry is spontaneously broken. Given identical sampling statistics, the statistical uncertainties in the three observables are comparable. Consequently, the D-Wave approach achieves equivalent statistical precision with approximately one-third the number of samples.

\begin{table}[h]
    \centering
    \begin{tabular}{c|c|c}
         & Classical & D-Wave \\
        \hline
        $\tau_{int}$ & 873.6 & 184.7\\
        $|\phi|$ & 0.8292(11) & 0.8270(14)\\
        $\chi_{|\phi|}$ & 2.39(15) & 2.66(26)\\
        $B_3^{|\phi|}$ & -0.8126(88) & -0.794(11)      
    \end{tabular}
    \caption{The autocorrelation time~\cite{Wolff:2003sm} and statistical uncertainty of observables $|\phi|$, $\chi_{|\phi|}$ and $B_3^{|\phi|}$ at $\lambda=10$, $\kappa=0.5$ on $64 \times 64$ lattice. To ensure a fair comparison, we compare the results with the same number of measurements (100) for both methods. This results in a total of $93,000 \times V$ updates for the classical Metropolis algorithm and $31,000 \times V$ updates for the D-Wave-based Metropolis-Hastings method, where $V$ denotes the lattice volume.
}
    \label{tab:autocor}
\end{table}

For nonzero $\kappa$, the neighboring fields should be taken into account via the first term in Eq. (\ref{eq:local_S}). 
\begin{align}
\label{eq:local_S}
S_{local}(x)=&-\kappa \sum_{\mu}(\phi(x+\hat{\mu})+\phi(x-\hat{\mu}))\phi(x) \nonumber\\ &+\phi(x)^2+\lambda(\phi(x)^2-1)^2 \,.
\end{align}
Here, the sum of the nearest 4 neighbor fields $\sum_{\mu}(\phi(x+\hat{\mu}) + \phi(x-\hat{\mu}))$ is considered as a boundary condition. 
A change in the number of qubits alters the possible values that the sum of neighboring fields, which we refer to as the boundary condition, can take. In order to keep a high acceptance rate and achieve more accurate simulations at large $\kappa$, we need to generate the distribution for each boundary condition. In Table~\ref{tab:one-site_bc}, we provide the number of boundary conditions in the case of Method III. However, due to limitations in allocated quantum computing resources, we do not present results using the distributions for different boundary conditions at nonzero $\kappa$.
\begin{table}[h]
    \centering
    \begin{tabular}{c|c|c}
       qubits  & size of QUBO & number of b.c. \\
       \hline
        1+4 & (11,11) & 121 \\
        1+5 & (16,16) & 249 \\
        1+6 & (22,22) & 505 \\
        1+7 & (29,29) & 1017 \\
    \end{tabular}
    \caption{The list of the QUBO matrix size and the number of distinct boundary conditions for single scalar field with one qubit for sign and other qubits for precision $n_{qubit}=4,5,6,7$.}
    \label{tab:one-site_bc}
\end{table}
The simulation procedures for $\kappa \neq 0$ are as follows:
\begin{enumerate}
    \item Initialize the field configuration.
    
    \item Update even lattice sites:
    \begin{enumerate}
        \item  Construct the QUBO for even sites using nearest-neighbor interactions and apply the appropriate boundary conditions.
        \item For each distinct boundary condition, generate samples using the D-Wave quantum annealer and store the resulting distributions.
        \item From the stored distribution, sample a candidate field configuration as a proposed update.
        \item Compute the Metropolis-Hastings acceptance probability.
        \item Accept or reject the proposed configuration based on the computed probability. If accepted, update the field at even sites.
    \end{enumerate}
    \item Update odd lattice sites: 
     Repeat steps 2(a)-2(e) for odd sites.
    \item Store the updated field configuration.
    \item Repeat from step 2 for the desired number of iterations.
\end{enumerate}

When $\kappa=0$, the field variables are completely decoupled due to the absence of nearest-neighbor interactions. As a result, the target distribution does not depend on the boundary conditions, and a single sampling is sufficient to represent all cases.

\begin{figure}
    \centering
    \includegraphics[width=\linewidth]{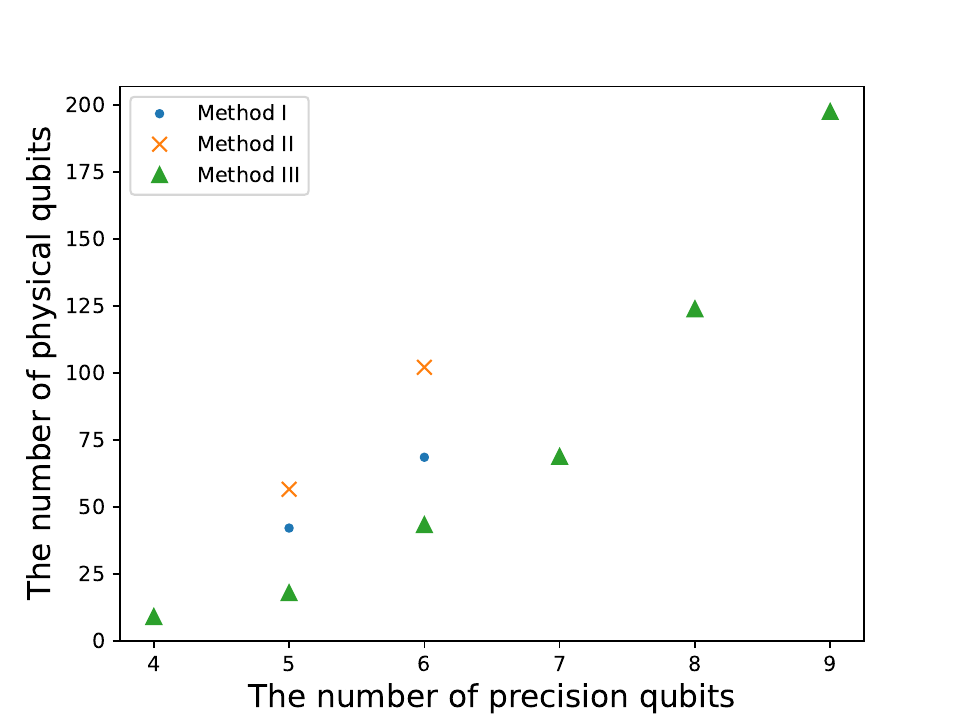}
    \caption{Scalability of physical qubits for the same precision embeddings when $\kappa=0$. The auxiliary qubits are not taken into account for the number of precision qubits.}
    \label{fig:scale}
\end{figure}

The scalability of the embeddings is shown in Fig.\ref{fig:scale}. Among the three methods, Method III requires the smallest number of physical qubits for a given precision. As the number of qubits used to represent the precision increases, the number of physical qubits required for the embedding grows rapidly, indicating an approximately exponential scaling behavior.

For three methods, the required number of logical qubits for the same precision is provided in Table~\ref{tab:logical}. 
\begin{table}[h]
    \centering
    \begin{tabular}{l|c|c}
       Method & precision qubits  & logical qubits \\
       \hline
        Method I   & $n_q=5,6$           & $n_q(n_q+1)/2$ \\
        Method II  & $n_q=5,6$           & $n_q^2$     \\
        Method III & 1+$n_q=4,5,6,7,8,9$ & $1+n_q(n_q+1)/2$ \\
    \end{tabular}
    \caption{The list of logical qubits in terms of $n_q$ for each method. For method III, precision qubits consist of one qubit for sign and qubits for resolution.}
    \label{tab:logical}
\end{table}

\section{Conclusion \label{sec:conclusion}}
In this study, we have investigated the implementation of two-dimensional Euclidean scalar $\phi^4$ field theory on a quantum annealer, focusing on the hybrid classical-quantum sampling approach. To accommodate the quartic interaction inherent in the $\phi^4$ potential, we developed and benchmarked three polynomial reduction techniques that transform higher order terms into quadratic QUBO representations suitable for the D-Wave Advantage2 system.

We demonstrated that, by employing these polynomial reductions along with appropriate auxiliary variables and optimized penalty functions, it is possible to sample scalar field configurations on the quantum annealer with a high validity and acceptance rate. Notably, Method III, which incorporates an explicit sign qubit, achieves significantly improved sampling fidelity while maintaining a lower physical qubit overhead compared to symmetry preserving but more resource-intensive alternatives.

Using the annealer-generated field distributions as a proposal mechanism in the Metropolis-Hastings algorithm, we performed large volume lattice simulations and validated our results against classical Metropolis updates. 
We observed good agreement in observables such as the expectation value ⟨$|\phi|$⟩, which allows us to confirm the expected critical behavior of the $\phi^4$ theory on a lattice. Notably, our hybrid algorithm maintained high acceptance rates even on larger lattices, particularly in the weak coupling regime.

Our results demonstrate that quantum annealers, despite current hardware limitations in connectivity and noise, can efficiently generate physically relevant configurations of scalar field theories. This work highlights the feasibility of using annealing-based devices for quantum field theoretic simulations and provides a scalable framework that can be extended to more complex theories and improved with next generation quantum annealing hardware.

Although the scalar field was digitized using only six qubits per site, we found no significant loss of precision compared to classical simulations performed with continuous variables. By utilizing histogram-based sampling generated on the D-Wave annealer, our hybrid algorithm achieved significantly shorter thermalization and autocorrelation times compared to conventional methods.

While it is in principle possible to perform similar histogram-based simulations on classical hardware by digitizing the scalar field, this approach is not feasible for simulations using continuous (real) variables. Moreover, in models without quartic interactions, where no auxiliary qubits are required, the available qubit budget can be devoted entirely to increasing the precision of the digitization. For example, when using 50 qubits per site to represent the scalar field, the quantum annealer can explore an extremely fine-grained field space. In contrast, performing equivalent histogram-based sampling on a classical computer would require enumerating $2^{50}$ field configurations per site, which is computationally infeasible due to the exponential scaling of memory and processing time.

While this work focused on the $\kappa=0$ case for quantum sampling, we demonstrated that the resulting distributions can still be used to efficiently simulate the $\kappa \neq 0$ regime via classical post-processing, achieving better performance than fully classical simulations. Furthermore, we proposed a method for extending quantum sampling directly to nonzero $\kappa$, which could offer a promising path toward even higher efficiency in future simulations.

\begin{acknowledgments}
J.K. thanks Wolfgang Unger for very useful and informative discussions and proofreading. The authors gratefully acknowledge the J\"ulich Supercomputing Centre for funding this project by providing computing time on the D-Wave Advantage\texttrademark{} System JUPSI through the J\"ulich UNified Infrastructure for Quantum computing (JUNIQ). This research was supported by ‘Quantum Information Science R\&D Ecosystem Creation’ through the National Research Foundation of Korea(NRF) funded by the Korean government (Ministry of Science and ICT(MSIT))(No. 2020M3H3A1110365). H.K. is supported by the National Research Foundation of Korea (NRF) grant funded by the Korean government (MSIT) (2023R1A2C1006542).

\end{acknowledgments}

\appendix
\section{Statistics \label{app:stats}}
\begin{figure*}
    \centering
    \subfigure[Monte Carlo history]{
    \includegraphics[width=0.45\linewidth]{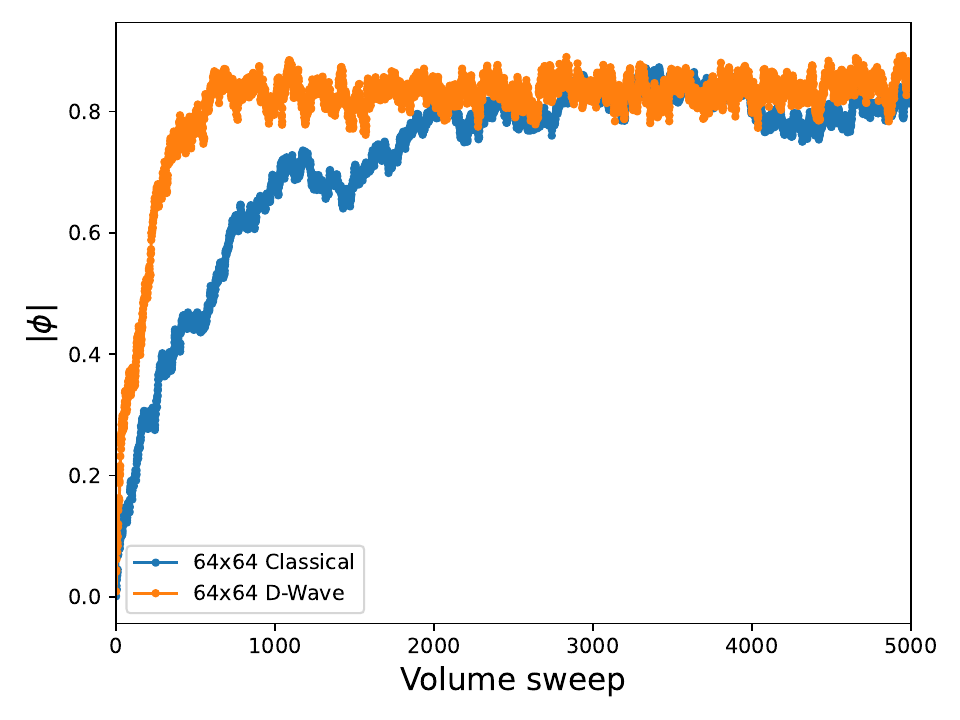}
    }
    \subfigure[Bin size analysis]{
    \includegraphics[width=0.45\linewidth]{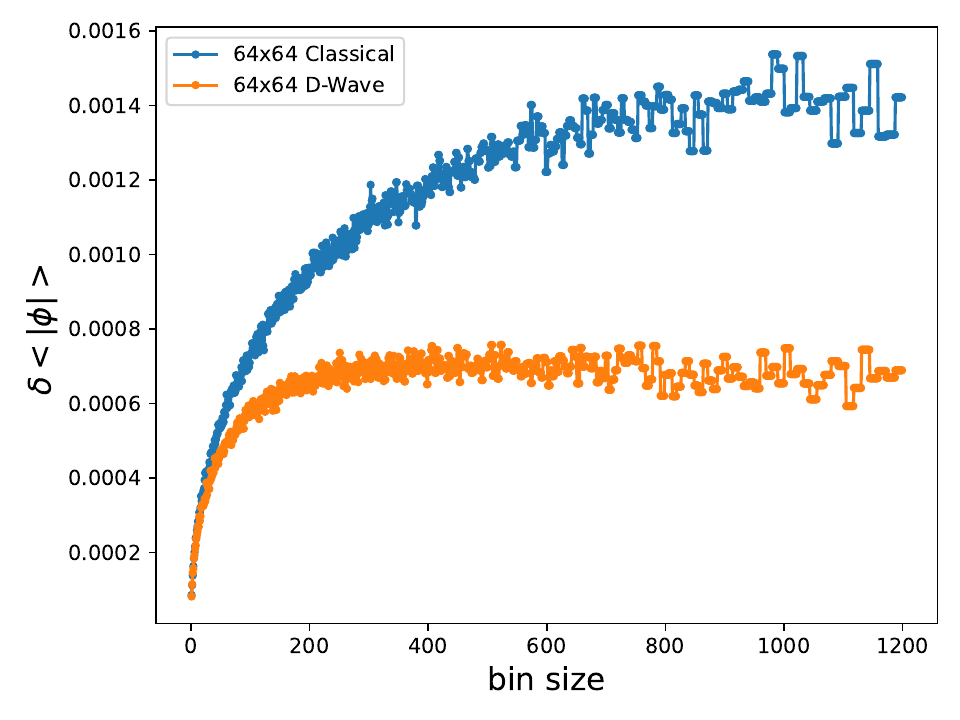}
    }
    \caption{(a): Monte Carlo history of Classical and D-Wave(method III) simulations. (b): The bin size is determined by requiring the statistical error of the observable $|\phi|$ to stabilize. This analysis is performed at $\lambda=10$, $\kappa=0.5$, on a $64 \times 64$ lattice.}
    \label{fig:MC}
\end{figure*}

We compare the autocorrelation times obtained using the classical Metropolis algorithm and the Metropolis-Hastings approach implemented on the D-Wave quantum annealer, as shown in Fig.\ref{fig:MC}. Based on this comparison, we set the number of thermalization sweeps to $3000$ for the classical algorithm and $1000$ for the D-Wave method. The bin sizes are chosen to be $900$ and $200$, respectively. For the data presented in Fig.\ref{fig:result_kappa}, Fig.\ref{fig:sus}, and Fig.\ref{fig:B3}, we perform $100,000 \times V$ update steps for both approaches.

\bibliography{references}

\end{document}